\documentclass[prl,twocolumn,superscriptaddress]{revtex4}

\usepackage{graphicx,t1enc}
\usepackage{epsfig}
\usepackage{epstopdf}
\usepackage{amssymb,amsmath}
\usepackage{color}
\usepackage{amstext}
\usepackage{amssymb}
\usepackage{graphicx}
\usepackage{esint}

\definecolor{blue}{rgb}{0,0,1}

\usepackage[inline]{trackchanges}

\begin{document}

\title[]{Enhancement of the flow of vibrated grains through narrow apertures by addition of small 
particles}

\author{Marcos A. Madrid}
\affiliation{Instituto de F\'isica de L\'iquidos y Sistemas Biol\'ogicos, CONICET, 59 789, 1900 La Plata, Argentina
and Departamento de Ingenier\'ia Mec\'anica, Universidad Tecnol\'ogica Nacional, Facultad Regional La Plata, La Plata, 1900, Argentina.}

\author{C. Manuel Carlevaro}
\affiliation{Instituto de F\'isica de L\'iquidos y Sistemas Biol\'ogicos, CONICET, 59 789, 1900 La Plata, Argentina
and Departamento de Ingenier\'ia Mec\'anica, Universidad Tecnol\'ogica Nacional, Facultad Regional La Plata, La Plata, 1900, Argentina.}

\author{Luis A. Pugnaloni}
\affiliation{Departamento de F\'isica, Facultad de Ciencias Exactas y Naturales, Universidad Nacional de La Pampa, CONICET,
Uruguay 151, 6300 Santa Rosa (La Pampa), Argentina.}

\author{Marcelo Kuperman}
\affiliation{Consejo Nacional de Investigaciones Cient\'{\i}ficas y T\'ecnicas, \\
Centro At\'omico Bariloche (CNEA), (8400) Bariloche, R\'{\i}o Negro, Argentina.}

\author{Sebasti\'an Bouzat}\thanks{email: bouzat@cab.cnea.gov.ar.}
\affiliation{Consejo Nacional de Investigaciones Cient\'{\i}ficas y T\'ecnicas, \\
Centro At\'omico Bariloche (CNEA), (8400) Bariloche, R\'{\i}o Negro, Argentina.}

\begin{abstract}
We analyze the flow and clogging of circular grains passing through a small aperture under vibration in two dimensions. Via Discrete Element Method simulations, we show that when grains smaller than the original ones are introduced in the system as an additive, the net flow of the original species can be significantly increased. Moreover, there is an optimal radius of the additive particles that maximizes the effect. This finding may constitute the basis for technological applications not only concerning the flow of granular materials but also regarding active matter, including pedestrian evacuation. 
\end{abstract}
\maketitle

\section{Introduction} The search for strategies to avoid clogs and enhance the flux of granular flows through narrow pathways is a problem of basic science that finds relevant 
applications in industry, as well as in crowd 
control in large scale facilities. The notable similarities found in the clogging 
dynamics of a wide variety of systems \cite{Zuriguel2014clogging} suggests a multiplicity of potential applications.

In recent years, various mechanisms to alleviate clogging of granular 
flows have been investigated. For instance, the effects 
of vibrating the setup \cite{mankoc2009role,lozano2012breaking}, making the exit 
oscillate \cite{to2017flow} and placing an obstacle in front of the exit 
\cite{zuriguel2011silo,endo2017obstacle} have been analyzed. The later strategy is able to reduce the clogging probability by two orders of magnitude even without applying vibrations \cite{zuriguel2011silo}. Whether these strategies can be extrapolated to other systems like crowds is still debated \cite{zuriguel2019contact}.  In Ref. \cite{nicolas2018} the authors called the attention to the fact that the 
evacuation of grains or pedestrians from a hopper-shaped enclosure with a small aperture may be 
sped-up by adding agents with a lower tendency to produce clogs than the original ones. Thus, mixing can also provide methods for alleviating clogging.


In this work, we study the flow of binary mixtures of grains of different sizes 
in a vibrated two-dimensional hopper by means of Discrete Element Method (DEM) simulations. The focus is set on studying the possibility of enhancing the effective flux of a given species (referred to as the {\em original} species) by introducing a relatively small amount of grains of smaller size (the {\em added} species). We perform exhaustive simulations that
shed light on the way the added particles should be chosen (or designed) in order to optimize the flow of the original ones. Our proposal is inspired by previous works that analyze the flow of mixtures of grains \cite{chevoir07,nicolas2018} and vibrated systems \cite{mankoc2009role}. Here we consider these ingredients together.

\begin{figure}
\includegraphics[width=0.5\columnwidth]{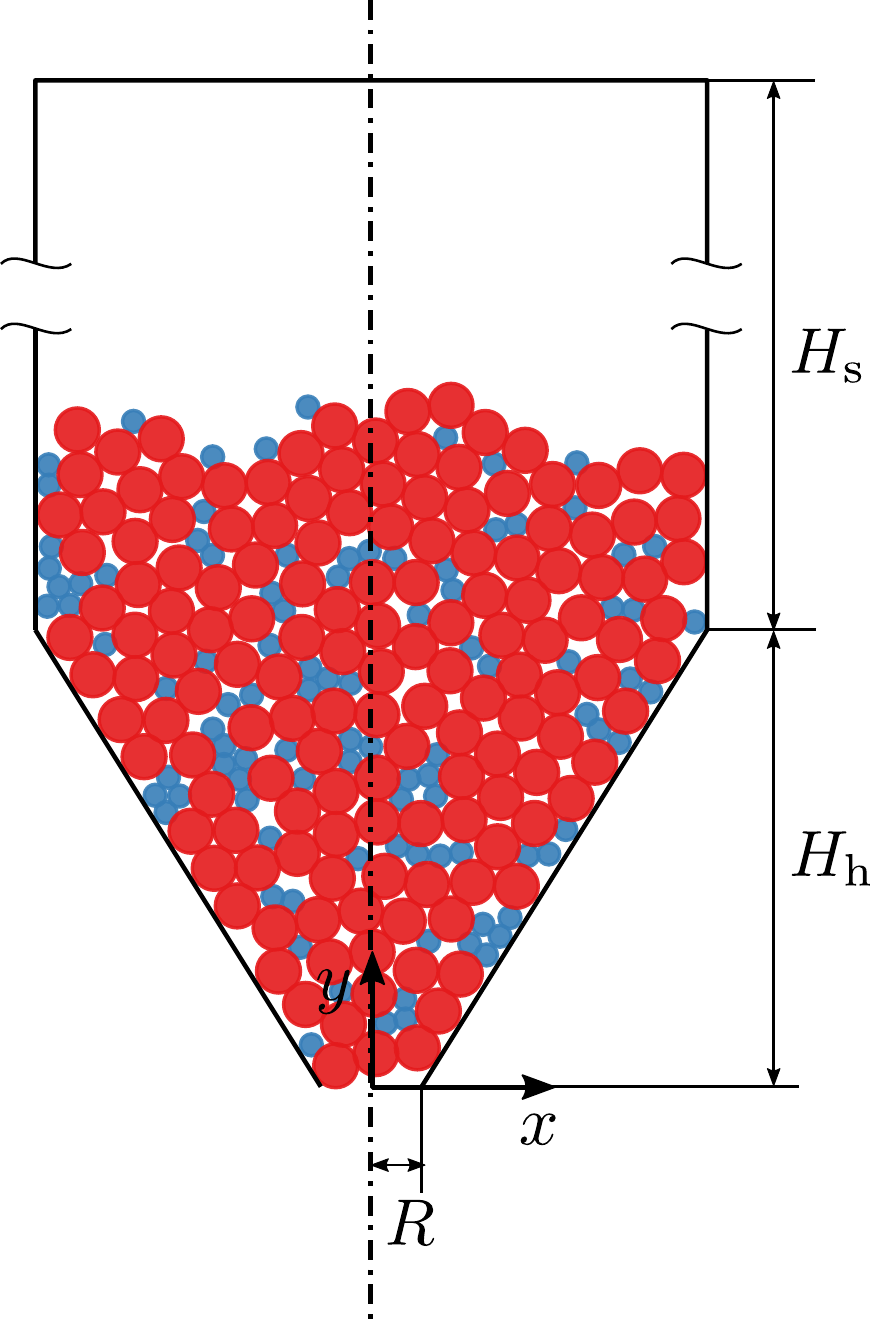}
\caption{\label{scheme} Sketch of the silo and hopper. The red circles correspond to the original species and the blue circles to the added particles.}
\end{figure} 

\section{Model and simulations} Our original grains are disks of radius $r_\text{o}=6.5$ cm and material density $1.0$ kg/m$^2$. The added species consists in disks of radius $r_\text{a}$ and same density as the main species, 
with $r_\text{o}/5 < r_\text{a} < r_\text{o}$. The total number of original and added grains are referred to as $N_\text{o}$ and $N_\text{a}$, respectively. The total number of particles $N_\text{o}+N_\text{a}=250$ is kept fixed in all our simulations.

We consider a vertical 2D silo and hopper of total height $H_\text{s} + H_\text{h} = 636$ cm $\approx 97.8 r_\text{o}$, width of $100$ cm $\simeq 15.3 r_\text{o}$ and aperture of radius $R=2.3 r_\text{o}$ (see Fig.~\ref{scheme}). The hopper height $H_\text{h} = 136$ cm leads to an angle $(\pi/6)$ between the hopper walls and the vertical direction. The acceleration of gravity is $g=9.8$ m/s$^2$.

Particles are initially deposited in the hopper randomly before opening its aperture. Each particle that crosses the aperture is re-injected above the granular column. Every $0.1$ s, an individual kick in random direction (uniformly distributed in [$0,2\pi$]) with an impulse uniformly distributed in the interval [$0,5\times10^{-5}$ Ns] is applied to each particle. At the same time, a global random kick with the same properties as the individual random kicks is applied to all the particles. The maximum duration of any clog is limited to $5$ s. If no particle flows through the aperture during $5$ s, our algorithm identifies each of the particles constituting the arch that blocks the aperture and removes these from the system and re-inject them at the top of the granular column. 


We used Box2D for the DEM simulations \cite{box2d,Catto}, which has been successfully used before to study granular flow \cite{Goldberg2015}, clogging \cite{Goldberg2018}, force networks \cite{Pugnaloni2016}, tapping \cite{Irastorza2013}, vigorous vibrations \cite{Sanchez2014}, stick-slip \cite{Carlevaro2020} and soil mechanics \cite{Pylos2015}. Particles are impenetrable (infinitely rigid). When at contact, the interaction is only defined by a restitution coefficient and a friction coefficient (static and dynamic friction are set equal). At each time step, Box2D uses a constraint solver based on Lagrange multipliers to calculate the normal contact forces that ensure that particles do not overlap~\cite{Pylos2015,bender2014}. Lagrange multipliers are also used to comply with the constraint of the  Coulomb criterion at each contact for the tangential contact force. After computing all the contact forces, the Newton-Euler equations of motion are integrated using a symplectic Euler algorithm \cite{bender2014}. In our simulations we set the restitution coefficient to a very low value ($0.1$) to reduce the number of time steps required for the system to come to rest when no external vibration is applied. The friction coefficient was set to $0.5$. The time step for solving the equations of motion was set to $1$ ms (simulations with shorter time steps yield the same results).

\section{Results} In our simulations we explore different particle size ratios $r=r_\text{a}/r_\text{o}$ and mixing ratios $\chi=N_\text{a}/(N_\text{a}+N_\text{o})$. For each set of parameters, we simulate $150$ s of the discharge and perform $50$ realizations with different initial random configurations. We compute the effective particle flow rate of the original particles $\tilde{Q}$ as the number of original particles discharged divided by the total time of the simulation. Here we use the tilde to distinguish from the total flow rate $Q$, which includes also the added particles. It is important to note that, strictly speaking, systems presenting clogging do not have a well defined flow rate. The flow rates $Q$ and $\tilde{Q}$ will depend on the cutoff used to break long lasting clogs. In our case, this cutoff is $5$ s. Despite this, the effective flow rate serves as a simple parameter to compare the flow between different systems when the cutoff used is fixed. 

\subsection{Enhancement of the flow rate by addition of small particles}

\begin{figure}
\centering
\includegraphics[width=.7\linewidth]{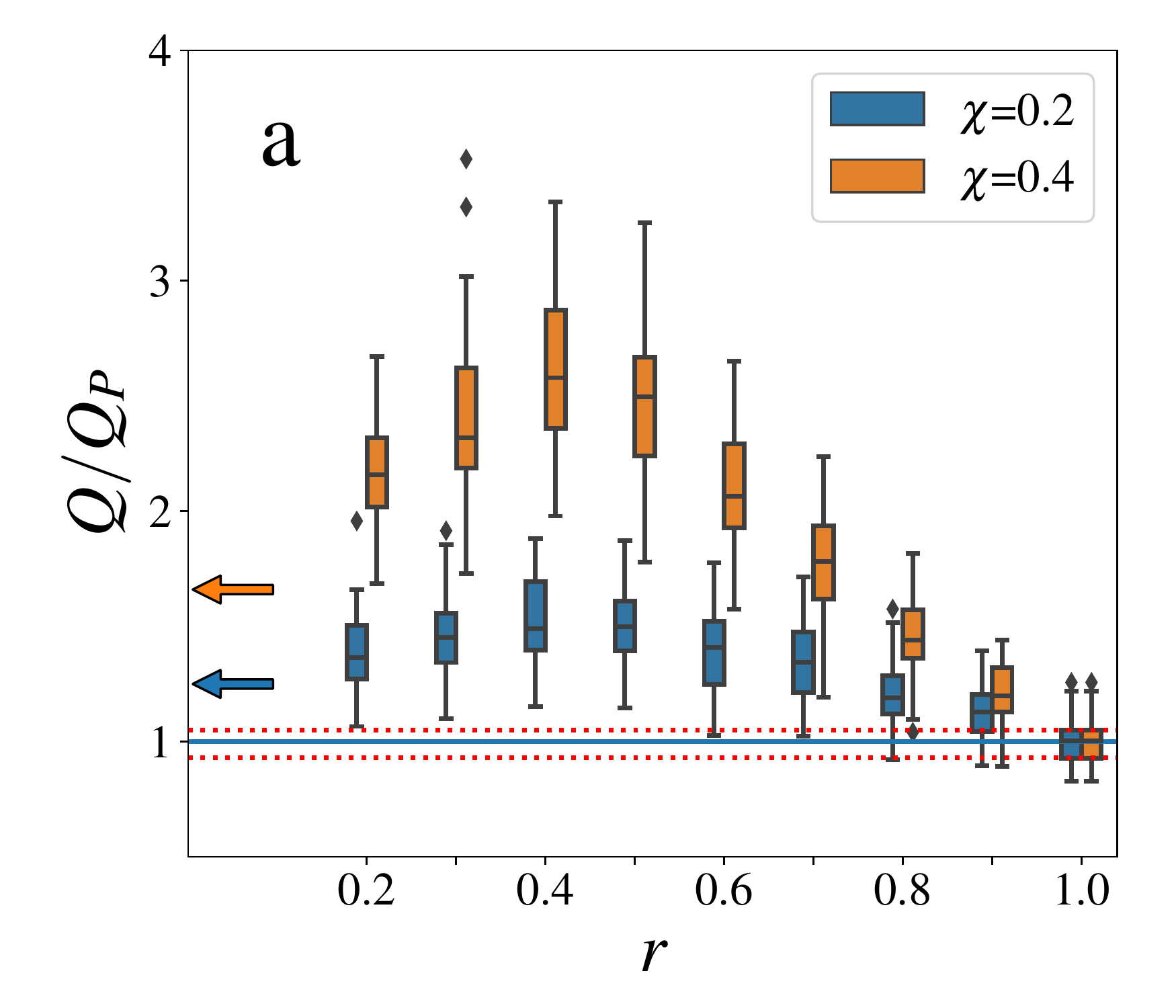} 
\vspace{-.1cm}
\includegraphics[width=.7\linewidth]{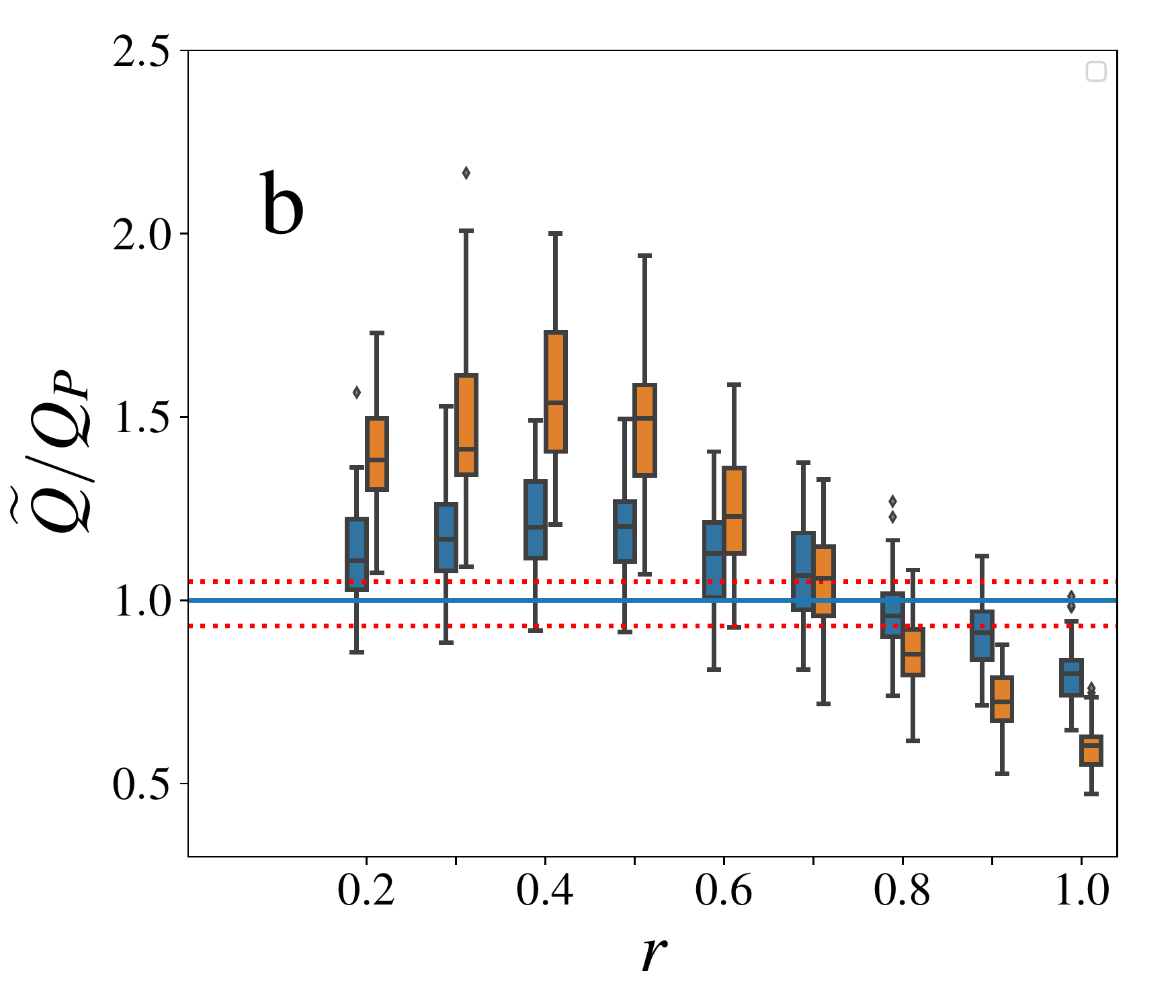} 
\caption{\label{fig2} Total particle flow rate $Q$ (a) and particle flow rate of the the original particles $\tilde{Q}$ (b) as a function of the size ratio $r$ for two mixing ratios $\chi=0.2$ (blue) and $0.4$ (orange) using added particles. The plotted quantities are divided by $Q_P$, which is the flow rate of the pure system (original particles alone). The box-plots indicate the quantiles for $0.065\%, 25\%, 50\%$ (median), $75\%$ and $99.35\%$, and rhombuses indicate outliers calculated over 50 realizations. The horizontal lines correspond to the quantiles for $25\%$ (bottom orange-dashed), $50\%$ (blue-solid), and $75\%$ (top orange-dashed) for a system of pure original particles. The arrows indicate the theoretical limiting value $Q=Q_P/(1-\chi)$ for $r \to 0$ (see text).}
\end{figure}

Figure \ref{fig2} shows results for $Q$ and for $\tilde{Q}$ as functions of $r$ for different values of $\chi$. The flow rates are scaled by the median of the flux of the {\em pure system} $Q_P\simeq 1.57 \text{ s}^{-1}$, 
that corresponds to the case $\chi=0$. 

Figure \ref{fig2}.a shows the total flow rate $Q$. In the limit $r\to 1$ we get the result of the pure system ($Q=Q_P$), since the two species become indistinguishable. Meanwhile, for $r\to 0$ (meaning arbitrarily small but finite sized particles) we expect the influence of the added species to be negligible and the clogging dynamics to be controlled essentially by the original particles. The total flow rate $Q$ would however be increased with respect to the pure system, since the tiny added grains flow together with the originals. In fact, by assuming that the evacuation of the $N_a+N_o$ particles in a mixed system with $r\to 0$ requires the same time as the evacuation of the $N_o$ particles in the pure system, a simple calculation leads to $Q=Q_P/(1-\chi)$. We are not able to run simulations for $r \to 0$ due to numerical instabilities when handling objects of very different sizes in Box2D \cite{Catto}; however, the results suggest a convergence to this analytic value (which is indicated with small arrows in Fig. \ref{fig2}.a). As Fig. \ref{fig2}.a shows, the dependence of $Q$ on $r$ between the limits $r\to 0$ and $r\to 1$ is far from being monotonic. Actually, $Q$ attains a maximum at $r\sim 0.4$ that is significantly larger than the $r \to 0 $ limit. 

In Fig. \ref{fig2}.b  we present results for the flow rate of the original particles $\tilde{Q}$ for the same simulations as in Fig. \ref{fig2}.a. Note that in the $r \to 1$ limit, for which the two species are indistinguishable (i.e., $Q=Q_P$, as discussed above), we expect a reduction in $\tilde{Q}$ with respect to $Q_p$, since a part of the flow of grains corresponds to the added particles. Namely, we have $\tilde{Q}(r=1)=(1-\chi)Q_P$. Meanwhile, for $r \to 0$, assuming that the added grains flow without affecting the flux of the original particles, we get $\tilde{Q}=Q_P$. The most striking feature in Fig. \ref{fig2}.b is that the values of $\tilde{Q}$ for the mixed systems exceed considerably the flow rate of the pure system for a wide range of the size of the added particles ($r<0.8$). Moreover, there exists an optimum size ratio close to $r=0.4$. For $\chi=0.4$, at the optimal $r$, $\tilde{Q}$ is about 50\% higher than $Q_P$. These results are particularly relevant since they indicate that the net flow of the original species can be significantly improved by adding smaller particles. This effect opens the way to important technological applications. Small particles could be used as additive to ease the flow though narrow constrictions. Interestingly, the optimum size of these particles is well below the size of the original particles, which means that screening with standard sieves can be used to remove the additive in a later part of the industrial process.

\begin{figure}[t]
 \includegraphics[width=\columnwidth]{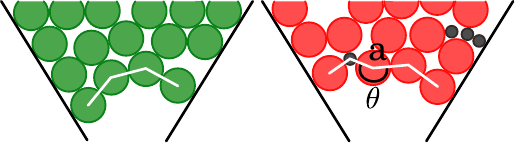}
 \caption{Snapshots from a simulations with original particles (left) and with added particles of $r=0.4$ (right). These are long-lasting arches (more than $5$ s) that needed to be removed. The branch vectors of the particles forming each arch are indicated with white segments. The large particle (indicated as ``a'') presents branch vectors with angle $\theta > 180$ Degree due to the neiboring small particle. This makes it especially unstable against vibrations \cite{lozano2012breaking}.}
 \label{fig:snapshot}
\end{figure}

To understand the ``lubricating'' effect of the presence of the added particles, one can focus attention on the arches that intermittently clog the exit \cite{duran98,garcimartin10,zuriguel05,to01}. Previous works showed that as the relative size of the exit in terms of the particle size increases, the clogging probability decreases \cite{to01,zuriguel05}. Hence, the addition of small particles reduces the mean size of the grains and this reduces clogging (the formation of an arch) overall. However, as we mentioned above, the effect in the mixed system is not monotonic as in a system of mono-sized particles whose radii are decreased. Moreover, one has to consider the destabilization of formed arches caused by vibration. For identical particles, arches were shown to break during vibration at the weakest bond, which is the particle for which its branch vectors form the maximum angle \cite{lozano2012breaking}. The higher the angle of the weakest bond, the more unstable the arch is. In our simulations, a small particle that participates in an arch tends to make one of its large neighbors to have a large angle between its branch vectors (see Fig. \ref{fig:snapshot}). These arches that contain a small particle are therefore less stable and tend to last shorter. As a consequence, the unclogging is enhanced. In summary, addition of  particles smaller than the originals not only reduce the arch formation (clogging) but also increases the arch destabilization (unclogging), leading to a notable increase in the flow rate. However, the effect is reduced for
$r\to 0$ because very small added particles have a low probability of cooperating with the large particles to form arches. Thus, in such a limit the clogging dynamics is dominated by the large grains, as mentioned before. Clearly, the reduction of the effect for decreasing $r$ leads to the existence of an optimal intermediate size of added particles that maximizes the flow rate.

The movie provided in the supplementary material \cite{supplementary1} helps to get intuition on the dynamics of the mixed system and allows to observe details of particular events close to the exit involving both large and small particles. In particular, it is possible to see that most long-lasting arches do not contain small particles.

\subsection{Long lasting clogs and survival probabilities}

\begin{figure}
\centering
\includegraphics[width=.80\linewidth]{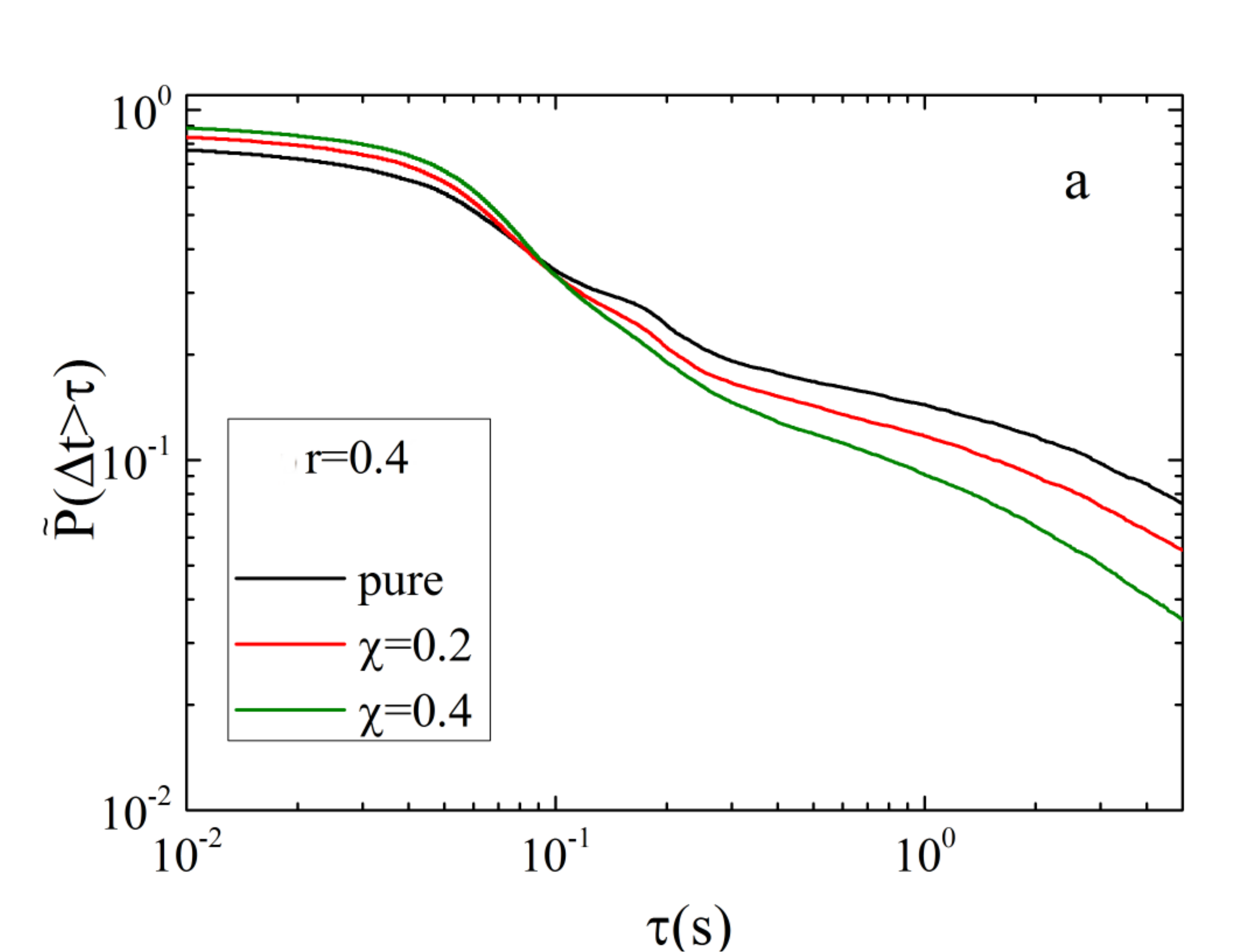}\\
\includegraphics[width=.80\linewidth]{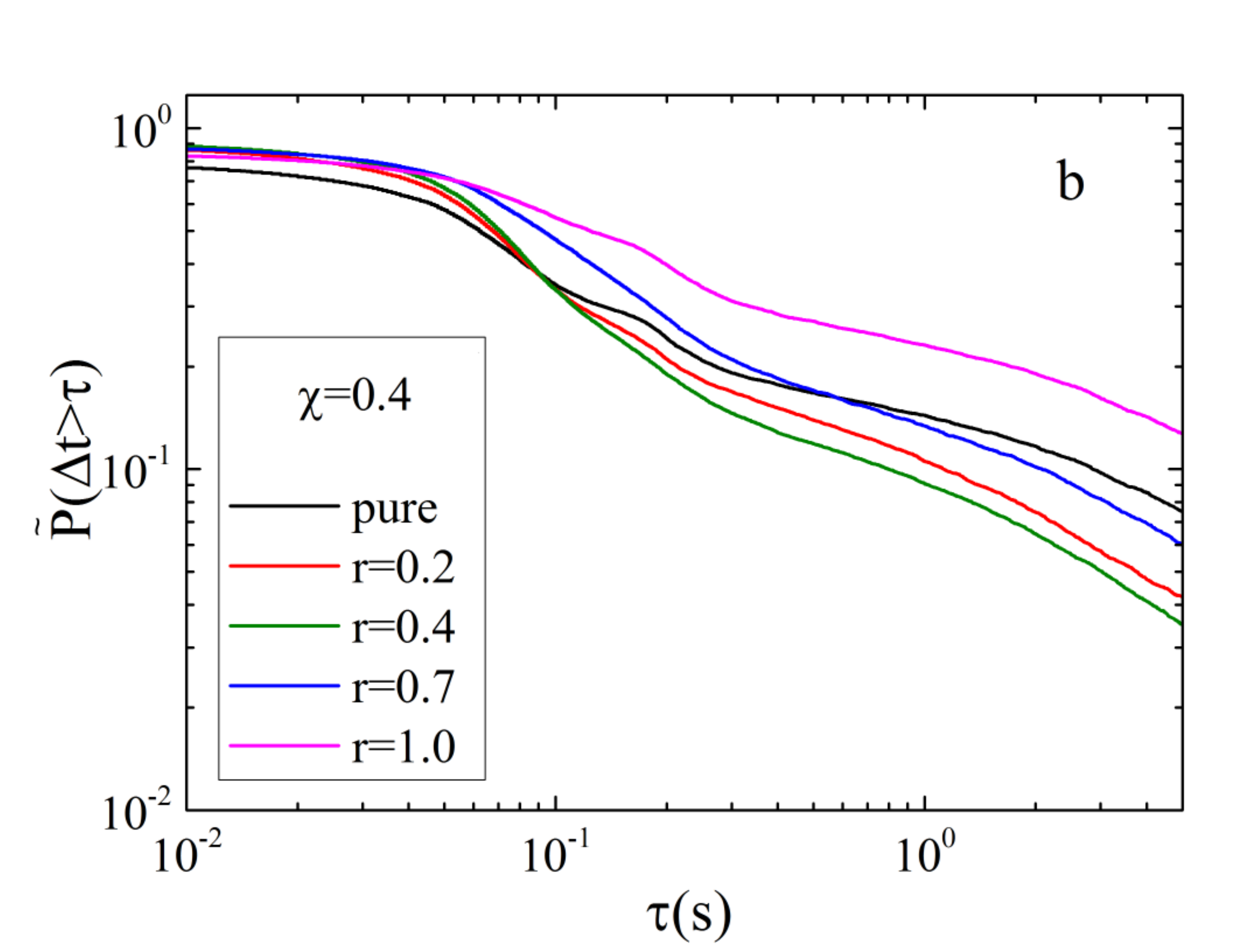}
\caption{\label{figsur} Survival function for different values of $\chi$ (a) and  $r$ (b). In both panels the black solid line corresponds to the pure system with no added particles.}
\end{figure}

To quantify the clogging dynamics we calculate the survival function $\tilde{P}(\Delta t\geq\tau)$, which gives the 
probability that we find a time gap $\Delta t > \tau$ between the egresses of two consecutive original particles. Note that a number of added particles may exit the container between the passage of two original particles. In Fig. \ref{figsur}.a we compare the survival function for the pure system with those for mixed 
systems with $\chi=0.2$ and $\chi=0.4$ considering a size ratio $r=0.4$. It is apparent that the likelihood of long lasting clogs decreases with $\chi$. This is the main driving for the enhancement 
of $\tilde{Q}$ with $\chi$ shown in Fig. \ref{fig2}.b. Note that 
due to the mechanism introduced to avoid clogs lasting longer than $5$ s, $\tilde{P}(\Delta t\geq\tau)$ is zero for $\Delta t>5$ s.

In Fig. \ref{figsur}.b we show the survival function for systems with varying $r$ at fixed $\chi=0.4$, whose flow rate are depicted in Fig. \ref{fig2}.b. We also plot the results for the pure system ($\chi=0$). 
Among the cases studied, the shortest tail of $\tilde{P}(\Delta t\geq\tau)$ is obtained for $r=0.4$, in concordance with the maximal flow of original particles. 

\begin{figure}[t]
\includegraphics[width=0.7\columnwidth]{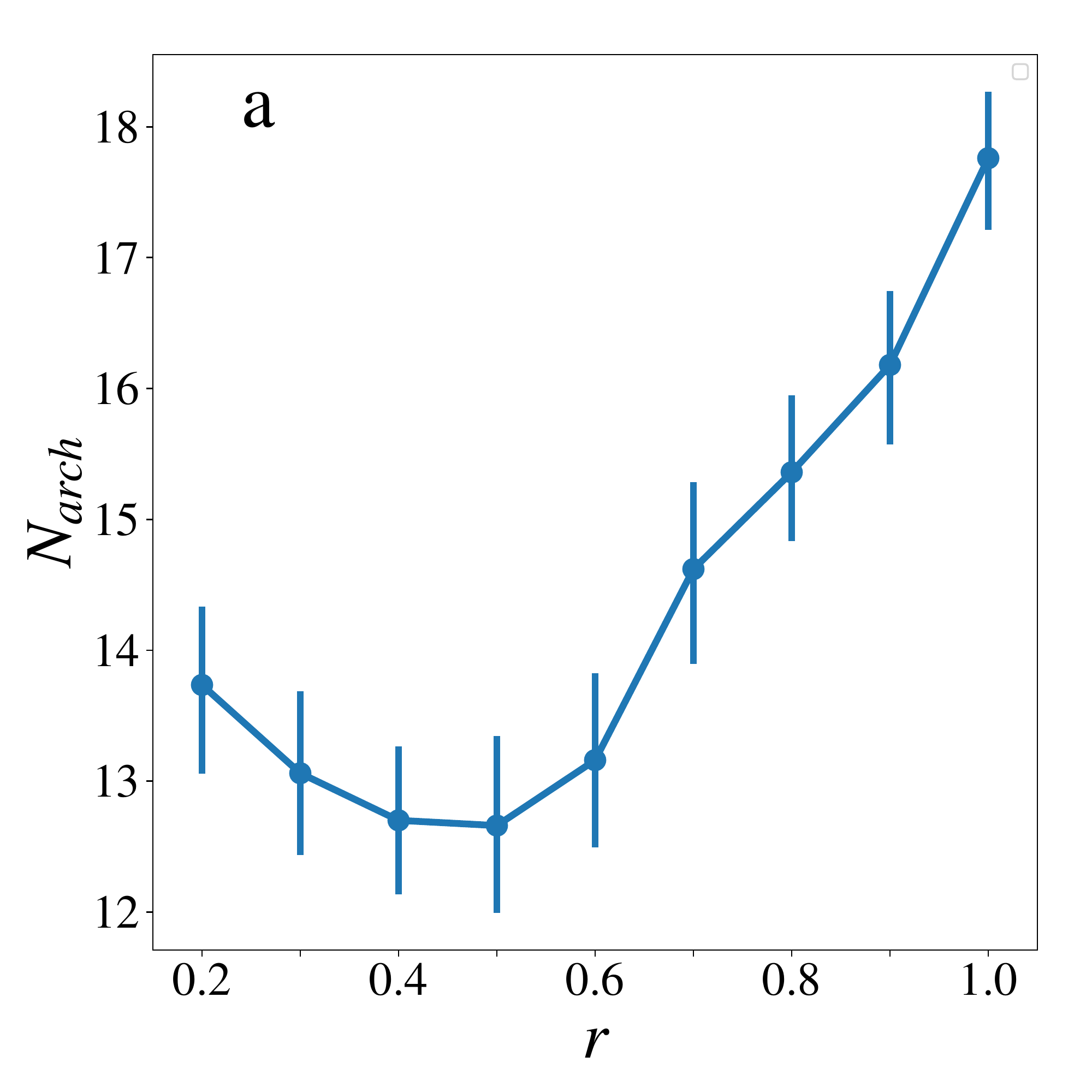}\\
\includegraphics[width=0.7\columnwidth]{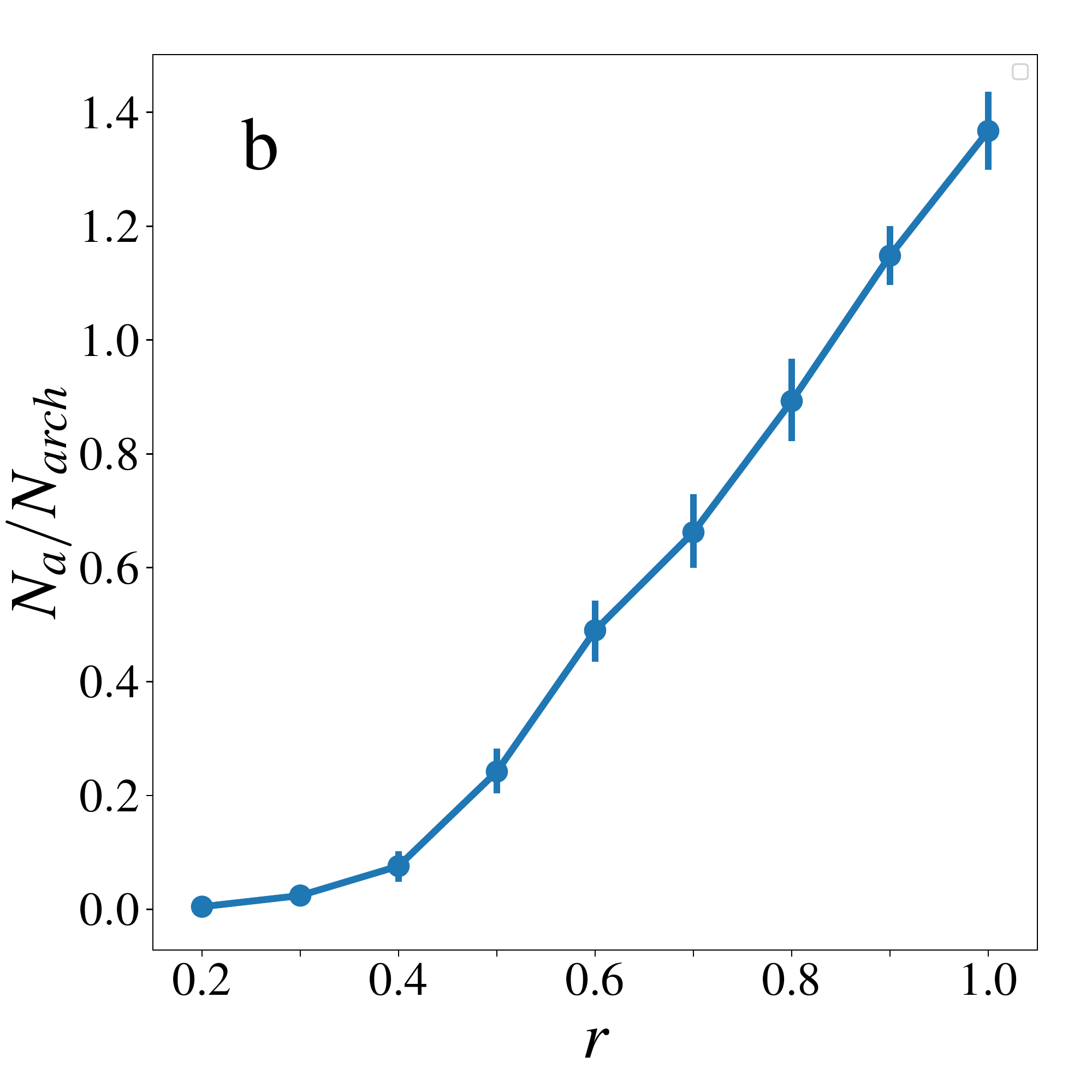}
\caption{\label{fig5} (a) Number $N_{\rm arch}$ of long lasting arches that block the aperture as a function of $r$ for $\chi=0.4$. Long lasting means $\tau >5$ s. The error bars correspond to the standard deviation over $50$ realizations. (b) Number $N_a$ of small added particles per arch in long lasting arches ($\tau >5$ s) as a function of $r$ for $\chi=0.4$.}
\end{figure}

As we mentioned, whenever a blocking arch lasts more than $5$ s we remove all the particles that form that arch to resume the flow. We have analyzed the composition of these long lasting arches. Figure \ref{fig5}.a shows the total number of long lasting arches ($\tau > 5$ s) that needed to be removed during any given simulation as a function of the radius $r$ of the added grains. Note that this number depends on the total simulated time. However, we have used the same simulation time in all cases and use this plot for a qualitative discussion about the effect on long-lasting arches. As we can see, the smaller the added particles the lower the occurrence of these long lasting arches. A subtle minimum is also seen around $r\approx 0.4$, in agreement with the observation of the maximum effective flow rate (see Fig.\ref{fig2}). 

The number of small particles involved in long lasting arches is shown in Fig. \ref{fig5}.b. The smaller the added particles, the less they participate in long lasting arches. This is because arches with small particles are less stable and rarely last beyond $5$ s. 

\section{Summary and conclusions \label{sec:conc}} 
We have shown that the clogging dynamics of a system of disks that flow through a narrow aperture under the effect of vibrations can be altered by the addition of a second species of smaller particles. In particular, the incidence of long lasting clogs is most reduced when the added particles are close to $0.4$ times the diameter of the original particles (for mixing ratios in the range $0.2 - 0.4$). Surprisingly, not only the total flow rate, but the effective flow rate of the original particles can be enhanced significantly. This provides a very effective and inexpensive mechanism to enhance the flow in clogging systems. Besides, since the added particles can be significantly smaller than the originals, they can be later removed using standard sieving techniques.

The effective use of the ``lubricating'' mechanisms described here to ease the flow in clogging scenarios may require the study of additional variables such as the system geometry (dimensionality, hopper angle, etc.), the particle-particle friction, the actual vibration amplitude and frequency, etc. Since a number of systems, very diverse in nature, present similar clogging dynamics, we expect that our finding can be valuable for systems as disparate as suspensions, bacteria colonies, granular materials and pedestrians.


\begin{thebibliography}{99}

\bibitem{Zuriguel2014clogging}
I.~Zuriguel, D.~R. Parisi, R.~C. Hidalgo, C.~Lozano, A.~Janda, P.~A. Gago,
  J.~P. Peralta, L.~M. Ferrer, L.~A. Pugnaloni, E.~Cl{\'e}ment, {\em et~al.},
  Clogging transition of many-particle systems flowing through bottlenecks,
  Sci. Rep. \textbf{4}, 7324 (2014).

\bibitem{mankoc2009role}
C.~Mankoc, A.~Garcimart{\'\i}n, I.~Zuriguel, D.~Maza, and L.~A. Pugnaloni,
  Role of vibrations in the jamming and unjamming of grains discharging from
  a silo, Phys. Rev. E \textbf{80}(1), 011309 (2009). 

\bibitem{lozano2012breaking}
C.~Lozano, G.~Lumay, I.~Zuriguel, R.~Hidalgo, and A.~Garcimart{\'\i}n,
  Breaking arches with vibrations: the role of defects, Phys. Rev. Lett. \textbf{109}(6), 068001 (2012).

\bibitem{to2017flow}
K.~To and H.-T. Tai, Flow and clog in a silo with oscillating exit, 
  Phys. Rev. E \textbf{96}(3), 032906 (2017).

\bibitem{zuriguel2011silo}
I.~Zuriguel, A.~Janda, A.~Garcimart{\'\i}n, C.~Lozano, R.~Ar{\'e}valo, and
  D.~Maza, Silo clogging reduction by the presence of an obstacle, Phys. Rev. Lett. \textbf{107}(27), 278001 (2011).

\bibitem{endo2017obstacle}
K.~Endo, K.~A. Reddy, and H.~Katsuragi, Obstacle-shape effect in a
  two-dimensional granular silo flow field, Phys. Rev. Fluids,
  \textbf{2}(9), 094302 (2017).

  
\bibitem{zuriguel2019contact}
I.~Zuriguel, I.~Echeverria, D.~Maza, R.~C.~Hidalgo, C.~Mart\'{i}n-G\'{o}mez, and A.~Garcimart\'{i}n, Contact forces and dynamics of pedestrians evacuating a room: the column effect, Safety Sci. \textbf{121}, 394 (2020).

\bibitem{nicolas2018} A. Nicolas, S. Ib\'a\~nez, M. N. Kuperman and S. Bouzat,
A counterintuitive way to speed up pedestrian and granular bottleneck flows prone to clogging: can 'more'
escape faster?, J. Stat. Mech. \textbf{2018}(8), 083403 (2018).

\bibitem{chevoir07} F. Chevoir, F. Gaulard and N. Roussel, Flow and jamming of granular mixtures through obstacles, Europhys. Lett. \textbf{79}(1), 14001 (2007).


  





\bibitem{box2d} Box2d physics engine, {https://www.box2d.org}

\bibitem{Catto} {E. Catto}, {Iterative dynamics with temporal coherence}, {(2005)},
{https://box2d.org/publications/}

\bibitem{Goldberg2015} E. Goldberg, C. M. Carlevaro, and L. A. Pugnaloni, Flow rate of polygonal grains through a bottleneck: Interplay between shape and size, {Pap. Phys.} \textbf{7}, 070016 (2015).

\bibitem{Goldberg2018} E. Goldberg, C. M. Carlevaro, and L. A. Pugnaloni, Clogging in two-dimensions: effect of particle shape, J. Stat. Mech. \textbf{11}, 113201 (2018).

\bibitem{Pugnaloni2016} L. A. Pugnaloni, C. M. Carlevaro, M. Kram\'ar, K. Mischaikow, and L. Kondic, Structure of force networks in tapped particulate systems of disks and pentagons. I. Clusters and loops, Phys. Rev. E \textbf{93}, 062902 (2016).

\bibitem{Irastorza2013} R. M. Irastorza, C. M. Carlevaro, and L. A. Pugnaloni, Exact predictions from the Edwards ensemble versus realistic simulations of tapped narrow two-dimensional granular columns, J. Stat. Mech. \textbf{2013}(12),  P12012 (2013).

\bibitem{Sanchez2014} M. S\'anchez, C. M. Carlevaro, and L. A. Pugnaloni, Effect of particle shape and fragmentation on the response of particle dampers, J. Vib. Control \textbf{20}(12), 1846 (2014).

\bibitem{Carlevaro2020} C. M. Carlevaro, R. Kozlowski, L. A. Pugnaloni, H. Zheng, J. E. S. Socolar, and L. Kondic,  Intruder in a two-dimensional granular system: Effects of dynamic and static basal friction on stick-slip and clogging dynamics, Phys. Rev. E \textbf{101}(1), 012909 (2020).

\bibitem{Pylos2015} M. Pytlos, M. Gilbert, and C. C. Smith, Modelling granular soil behaviour using a physics engine, Géotechnique Lett.
\textbf{5}(4), 243 (2015).

\bibitem{bender2014} J. Bender, K. Erleben, J. Trinkle, Interactive Simulation of Rigid Body Dynamics in Computer Graphics, Comput. Graph. Forum \textbf{33}(1), 246 (2014).

\bibitem{duran98}
J. Duran, E. Kolb, and L. Vanel, Static friction and arch formation in granular materials, Phys. Rev. E \textbf{58}, 805 (1998).

\bibitem{garcimartin10}A. Garcimart\'in, I. Zuriguel, L. A. Pugnaloni, and A. Janda, Shape of jamming arches in two-dimensional deposits of granular materials, Phys. Rev. E \textbf{82}, 031306 (2010).

\bibitem{zuriguel05} I. Zuriguel, A. Garcimart\'in, D. Maza, L. A. Pugnaloni, and J. M. Pastor, Jamming during the discharge of granular matter from a silo, Phys. Rev. E \textbf{71}, 051303 (2005).


\bibitem{to01}K To, P Lai, and H. K. Pak, Jamming of Granular Flow in a Two-Dimensional Hopper, Phys. Rev. Lett. \textbf{86}, 71 (2001).

\bibitem{supplementary1} See supplementary material: Time-lapse movie of simulations showing, side by side, the discharge of the pure system and the system with added grains of small size.


 
\end{thebibliography}
\end{document}